\documentstyle[prl,aps,preprint,floats]{revtex}

\begin{document}

\title{Generalized plane-fronted gravitational waves in any dimension}

\author{Yuri N.~Obukhov\footnote{On leave from: Dept. of Theoret. 
Physics, Moscow State University, 117234 Moscow, Russia}}
\address{Institute for Theoretical Physics, University of Cologne,
50923 K\"oln, Germany}

\maketitle

\begin{abstract}
We study the gravitational waves in spacetimes of arbitrary dimension.
They generalize the $pp$-waves and the Kundt waves, obtained earlier 
in four dimensions. Explicit solutions of the Einstein and Einstein-Maxwell
equations are derived for an arbitrary cosmological constant. 
\end{abstract}
\bigskip


\section{Introduction}

Recently, there has been some progress in deriving the various generalizations 
of the plane wave solutions of the Maxwell and Einstein-Maxwell equations 
in higher-dimensional spacetimes \cite{coley,hervik,gimon}. The interest in 
such configurations is motivated, in particular, by applications that the 
plane wave solutions find in string theory, see e.g. \cite{ark,str}
and the references therein. The gravitational plane-fronted waves in four 
dimensions represent a well known class of solutions which satisfy the
so-called radiative conditions \cite{peres,pen,griff,vdz,exact}. 

In addition, some attention has been paid recently to the generalizations 
of the plane wave solutions to metric-affine theories of gravity with
torsion and nonmetricity waves along with the usual gravitational 
waves \cite{babu,dirk1,dirk2,king}. 

The aim of this work is to present exact gravitational wave solutions
of the Einstein field equation with a cosmological term $\Lambda$ for
arbitrary dimensions $2 + N$. 

In Sec.~\ref{ansatz}, the geometric ansatz for the metric (coframe) is
formulated as an immediate generalization of the previous work done in four
dimensions \cite{pleb,orr}. We demonstrate that the resulting curvature
of spacetime satisfies the generalized radiation conditions. Using this
ansatz, in Sec.~\ref{EE} we derive a partial differential equation for the 
only unknown function. The explicit solutions are obtained in Sec.~\ref{PPW}.
Our attention is not only confined to higher dimensions. The wave solutions 
in two and three spacetime dimensions represent a particular case which is 
analyzed separately in Sec.~\ref{PPW}. In the absence of the Maxwell radiative
source, such spacetimes are isometric to the de Sitter (or anti-de Sitter) 
manifold. The four-dimensional case is only briefly mentioned. The 
higher-dimensional wave solutions are obtained under the assumption of 
the rotational symmetry. We study in detail the case of the positive 
cosmological constant $\Lambda >0$. The solutions for the negative values 
$\Lambda <0$ are briefly discussed in Sec.~\ref{Neg}. In Sec.~\ref{DC} we 
summarize the results obtained and the Appendix contains some useful 
mathematical definitions and computations.

\section{Geometry of the ansatz}\label{ansatz}

Let us denote the $2 + N$ local spacetime coordinates by $x^i = \{\sigma, 
\rho, z^1,\dots, z^N\}$, with $i=0,1,\dots,N+1$. The upper case Latin indices, 
$A,B,\dots = 0, 1$, label the first 2 spacetime dimensions which are relevant 
to a $pp$-wave. In particular, $x^A = \{\sigma, \rho\}$ are the wave 
coordinates with the wave fronts described by the surfaces of constant 
$\sigma$, and $\rho$ is an affine parameter along the wave vector of the null 
geodesic. The lower case Latin indices, $a,b,\dots = 1,\dots,N$, refer to 
an $N$-dimensional space of constant curvature. Greek indices, $\alpha, \beta,
\dots = 0,1,\dots,N+1$, label the local anholonomic (co)frame components. We 
denote separate frame components by a circumflex over the corresponding index 
in order to distinguish them from coordinate components. The differential of 
an arbitrary function $f(x^i)=f(\sigma,\rho,z^a)$ with respect to the last 
$N$ coordinates is denoted $\underline{d}f:= dz^a\,\partial_af$. 

The line element reads
\begin{equation}
ds^2 = g_{\alpha\beta}\,\vartheta^\alpha\otimes\vartheta^\beta,\label{ds2}
\end{equation}
with the half-null Lorentz metric 
\begin{equation}
g_{\alpha\beta} = \left(\begin{array}{cc}g_{AB}&0\\ 0&g_{ab}\end{array}
\right),\qquad g_{AB} = \left(\begin{array}{cc}0&1\\ 1&0\end{array}\right),
\quad g_{ab} = \delta_{ab}.\label{met}
\end{equation}
With the functions $p(z^a), q(\sigma, z^a), s(\sigma,\rho,z^a)$, the 
components of the coframe 1-form are given by
\begin{equation}
\vartheta^{\widehat{0}} = -\,d\sigma,\qquad \vartheta^{\widehat{1}} 
= \left({\frac q p}\right)^2\left(s\,d\sigma + d\rho\right), 
\qquad \vartheta^a = {\frac 1 p}\,dz^a,\quad a=1,\dots,N.\label{cof}
\end{equation}
The dual frame basis (such that $e_\alpha\rfloor\vartheta^\beta = 
\delta_\alpha^\beta$) reads:
\begin{equation}
e_{\widehat{0}} = -\,\partial_\sigma + s\,\partial_\rho,\qquad 
e_{\widehat{1}} = \left({\frac p q}\right)^2\partial_\rho,\qquad
e_a = p\,\partial_a. 
\end{equation}
We choose 
\begin{eqnarray}
p &=& 1 + {\frac \Lambda 4}\,z_a z^a,\qquad q = q_0\,\alpha + \beta_a
\,z^a,\qquad q_0 = 1 - {\frac \Lambda 4}\,z_a z^a,\label{pq}\\ 
s &=& -\,{\frac {\rho^2} 2}\,(\Lambda\alpha^2 + \beta_a\beta^a) 
+ \rho\,{\frac {\partial_\sigma q}q} + {\frac {p^{N/2}}{2\,q}}
\,H(\sigma, z^a).\label{pqs}
\end{eqnarray}
Here $\Lambda$ is constant, the $1+N$ functions $\alpha = \alpha(\sigma)$ 
and $\beta_a = \beta_a(\sigma)$ depend on the coordinate $\sigma$ only, 
and $z_a z^a = \delta_{ab}\,z^az^b$, $\beta_a\beta^a = \delta_{ab}\,\beta^a
\beta^b$; $H(\sigma, z^a)$ is an unknown function to be determined by the 
field equation. Then, for {\it arbitrary} $\alpha,\beta_a$, the curvature 
2-form reads:
\begin{equation}
R^{\alpha\beta} = -\,\Lambda\,\vartheta^\alpha\wedge\vartheta^\beta + 2
\,\gamma^{[\alpha}\,k^{\beta]}\wedge\vartheta^{\widehat{0}}.\label{curv}
\end{equation}
Here the null vector is $k_\alpha = \delta_{\alpha}^{\widehat{0}}$, with 
$k_\alpha k^\alpha = 0$, and the components of the co-vector-valued 1-form 
$\gamma_\alpha$ are given in terms of the derivatives of the function $s$:
\begin{equation}\label{gam}
\gamma_{\widehat{0}} = 0,\qquad\gamma_{\widehat{1}} = 0,\qquad 
\gamma_a = - \underline{D}\left[\left({\frac q p}\right)^2e_a\rfloor
\underline{d}s\right] + \left({\frac q p}\right)^3e_a\rfloor\left[
\underline{d}\left({\frac p q}\right)\wedge\underline{d}s\right]. 
\end{equation}
The $N$-dimensional covariant derivative is denoted as $\underline{D}$ (the
details of computations of the connection and the curvature are presented 
in the Appendix). The 1-form (\ref{gam}) ``lives" on the $N$-submanifold and
as a result, it has the evident properties
\begin{equation}
\vartheta^\alpha\wedge\gamma_\alpha = 0,\qquad e_A \rfloor\gamma_\alpha 
= 0,\qquad e_\alpha\rfloor\gamma^\alpha = e_a\rfloor\gamma^a. 
\end{equation}
Substituting (\ref{pq}), (\ref{pqs}) into (\ref{gam}), we find the trace 
of the 1-form explicitly:
\begin{equation}
e_a\rfloor\gamma^a = -\,{\frac {p^{N/2}q}2}\left(\partial^a\partial_a H 
+ {\frac {\Lambda N(N+2)}{4p^2}}\,H\right).\label{trace}
\end{equation}

In view of the obvious orthogonality relation $\gamma_\alpha k^\alpha = 0$,
the tensor-valued 2-form $S^{\alpha\beta} = R^{\alpha\beta} + \Lambda
\,\vartheta^\alpha\wedge\vartheta^\beta$ satisfies the so-called radiation
conditions \cite{orr}: 
\begin{equation}\label{rad}
S^{\alpha\beta}\,k_\beta = 0,\qquad S^{[\alpha\beta}\,k^{\gamma]} =0.
\end{equation}
The same radiation conditions are also fulfilled by the Weyl 2-form which 
can be derived from (\ref{curv}):
\begin{equation}
W^{\alpha\beta} = 2\,{\nearrow\!\!\!\!\!\!\!\gamma\ }^{[\alpha}\,k^{\beta]}
\wedge\vartheta^{\widehat{0}}.
\end{equation}
This formula involves the traceless part ${\nearrow\!\!\!\!\!\!\!\gamma
\ }^\alpha$ of the 1-form (\ref{gam}) which is defined by 
${\nearrow\!\!\!\!\!\!\!\gamma\ }^{\widehat{0}} ={\nearrow\!\!\!\!\!\!\!
\gamma\ }^{\widehat{1}} = 0$ and by ${\nearrow\!\!\!\!\!\!\!\gamma\ }^a 
:= \gamma^a - 1/N\,(e_b\rfloor\gamma^b)\,\vartheta^a$. 
By definition, we have $e_\alpha\rfloor{\nearrow\!\!\!\!\!\!\!\gamma
\ }^{\alpha} = e_a\rfloor {\nearrow\!\!\!\!\!\!\!\gamma\ }^a = 0$.

An arbitrary function $w = w(\sigma)$ defines the combined coordinate and 
coframe (Lorentz) transformation
\begin{equation}\label{trans1}
\sigma\longrightarrow\int^\sigma w^2(\sigma)d\sigma,\qquad
\vartheta^A\longrightarrow L^A{}_B\vartheta^B,\quad
L^A{}_B = \left(\begin{array}{cc}w^2 & 0\\ 0 & w^{-2}\end{array}\right),
\end{equation}
which leaves invariant the form of the line element but rescales the three
functions as follows:
\begin{equation}
\alpha\rightarrow\alpha/w,\qquad \beta_a\rightarrow\beta_a/w,\qquad
H\rightarrow H/w^2.\label{trans2}
\end{equation}
As a result, a nonzero $\alpha(\sigma)$ can always be rescaled to $\alpha =1$
with the help of (\ref{trans1}), (\ref{trans2}) by choosing $w = \alpha$.

\section{Gravitational equation}\label{EE}

The Ricci 1-form reads:
\begin{equation}
e_\beta\rfloor R^{\alpha\beta} = \Lambda (N + 1)\,\vartheta^\alpha -
(e_\alpha\rfloor\gamma^\alpha)\,k^\alpha\,\vartheta^{\widehat{0}} =
\left[\Lambda (N + 1)\,\delta^\alpha_\beta - (e_a\rfloor\gamma^a)
\,k^\alpha k_\beta\right]\vartheta^\beta. 
\end{equation}
Here we took into account that $\vartheta^{\widehat{0}} = k_\alpha
\vartheta^\alpha$. 

In the tensor language, the Ricci tensor thus has the form:
\begin{equation}
Ric_{\alpha\beta} = \Lambda (N + 1)\,g_{\alpha\beta} - (e_a\rfloor\gamma^a)
\,k_\alpha k_\beta.\label{ric}
\end{equation}
Accordingly, the metric (\ref{ds2})-(\ref{pqs}) is, by construction, 
a solution of the Einstein field equation with the cosmological term 
$\Lambda (N + 1)$ and the ``pure radiation" source. The latter is described 
by the null vector field $k_\alpha$ and the radiation energy density 
$-(e_a\rfloor\gamma^a)$. 

A particular example of such a material source is represented by the
energy-momentum of an electromagnetic wave. Taking the potential 1-form 
\begin{equation}
A = \varphi(\sigma,z^a)\,\vartheta^{\widehat{0}},
\end{equation}
we find the 2-form of the electromagnetic field $F = dA = {\frac 12}
\,F_{\alpha\beta}\vartheta^\alpha\wedge\vartheta^\beta$ with the tensor 
components 
\begin{equation}
F_{\alpha\beta} = 2n_{[\alpha}\,k_{\beta]}.\label{F}
\end{equation}
Here $n^\alpha\,k_\alpha =0$. In terms of the vector potential, the covector
$n$ reads as follows:  
\begin{equation}
n_{\widehat{0}} =0,\qquad n_{\widehat{1}} = 0,\qquad 
n_a = e_a\rfloor d\varphi = p\,\partial_a\varphi.
\end{equation}
The unknown scalar function $\varphi$ is determined by the Maxwell
equation $d\,{}^\ast\! F = 0$ that for the metric (\ref{ds2})-(\ref{cof}) 
reduces to the partial differential equation
\begin{equation}
\partial_a\left(p^{2-N}\,\partial^a\varphi\right) = 0.\label{ddf}
\end{equation}
We can see from (\ref{F}) that
\begin{equation}
F_{\alpha\gamma}\,F_\beta{}^\gamma = n_\gamma n^\gamma\,k_\alpha k_\beta,
\qquad F_{\alpha\beta}\,F^{\alpha\beta} =0,\label{FF}
\end{equation}
and hence the energy-momentum reads $T_{\alpha\beta} = c\,\varepsilon_0\left(
F_{\alpha\gamma}\,F_\beta{}^\gamma - {\frac 14}g_{\alpha\beta}\,F_{\rho\sigma}
F^{\rho\sigma}\right) = c\,\varepsilon_0 n_\gamma n^\gamma k_\alpha k_\beta$.
Combining (\ref{trace}), (\ref{ric}), and (\ref{FF}), we find that Einstein's 
equation 
\begin{equation}
Ric_{\alpha\beta}- {\frac 12}R\,g_{\alpha\beta} + \lambda\,g_{\alpha\beta}
= {\frac {8\pi G}{c^3}}\,T_{\alpha\beta}
\end{equation}
with cosmological constant $\lambda = \Lambda (N + 1)$ is satisfied provided
the unknown function $H(\sigma, z^a)$ is a solution of
\begin{equation}\label{ddH}
\partial^a\partial_a H + {\frac {\Lambda N(N+2)}{4p^2}}\,H = {\frac {16\pi 
G\varepsilon_0\,p^{2-N/2}}{c^2\,q}}\,\partial_a\varphi\partial^a\varphi. 
\end{equation}
Here $G$ is Newton's gravitational constant and $\varepsilon_0$ the electric 
constant of the vacuum (vacuum permittivity).

\section{Vacuum solution in any dimension}

In vacuum, when $\varphi =0$, one can verify that the partial differential 
equation (\ref{ddH}) has a solution:
\begin{equation}\label{H1}
H(\sigma, z^a) = f(\sigma)\,H_1(z^a),
\qquad H_1 = {\frac {\widetilde{q}} {p^{N/2}}}.
\end{equation}
Here $f(\sigma)$ is an arbitrary function and 
\begin{equation}
\widetilde{q} = \left(1 - {\frac \Lambda 4}\,z_a z^a\right)
\widetilde{\alpha} + \widetilde{\beta}_a\,z^a.\label{qt}
\end{equation}
This has the same structure as the function $q$ from (\ref{pq}), but the
arbitrary functions $\widetilde{\alpha} = \widetilde{\alpha}(\sigma)$ 
and $\widetilde{\beta}_a = \widetilde{\beta}_a(\sigma)$ are {\it different} 
from $\alpha,\beta_a$, in general.

Substituting this into (\ref{pqs}), we find 
\begin{equation}
s = -\,{\frac {\rho^2} 2}\,(\Lambda\alpha^2 + \beta_a\beta^a) + {\frac 
{\rho\,(\partial_\sigma q) + \widetilde{q}\,f(\sigma)/2}q}.\label{pqs0}
\end{equation}
As a result, we can verify that the 1-form (\ref{gam}) vanishes for such 
a solution, and the spacetime curvature (\ref{curv}) becomes constant:
$R^{\alpha\beta} = -\,\Lambda\,\vartheta^\alpha\wedge\vartheta^\beta$. 
The resulting metric 
\begin{equation}
ds^2 = -\,2\left({\frac qp}\right)^2\left(s\,d\sigma^2 
+ d\sigma d\rho\right) + {\frac {dz_adz^a}{p^2}}
\end{equation}
with (\ref{pqs0}) and (\ref{qt}) thus represents the $(2 + N)$-dimensional 
de Sitter (or anti-de Sitter, depending on the sign of $\Lambda$) spacetime 
with a gravitational $pp$-wave propagating in it. In four dimensions ($N=2$) 
this was demonstrated in \cite{pod}. 

The particular solution (\ref{H1}) satisfies the vacuum partial differential 
equation (\ref{ddH}) for any dimension $N$. The complete solution depends
significantly on the dimension, and it is instructive to analyze the 
different values of $N$ separately.

\section{Explicit solutions for positive cosmological constant}\label{PPW}

\subsection{$N=0$: Two-dimensional spacetime}

In the two-dimensional spacetime, the $z$ coordinates are absent. Hence 
the functions (\ref{pqs}) reduce to $p=1$, $q=\alpha(\sigma)$, and $s = 
(-\,\Lambda\rho^2\alpha^2 + H)/2$, with $H=H(\sigma)$. We do not 
have an equation (\ref{ddH}) in this case, and the resulting metric reads: 
\begin{equation}
ds^2 = -\,2\alpha^2\left({\frac {-\Lambda\rho^2\alpha^2 + H(\sigma)}
2}\,d\sigma^2 + d\sigma d\rho\right).\label{N0met}
\end{equation}
We can put $\alpha =1$ by means of the rescaling (\ref{trans1}), 
(\ref{trans2}).

\subsection{$N=1$: Three-dimensional spacetime}

In three spacetime dimensions, we have $N=1$ and a single $z$-coordinate, 
so that the partial derivatives in (\ref{ddH}) reduce to the ordinary ones. 
Then (\ref{H1}), (\ref{qt}) evidently describes a general solution of the 
resulting homogeneous equation, with $\widetilde{\alpha}$ and $\widetilde{
\beta}$ being the two arbitrary integration functions. We rewrite this 
general solution of the homogeneous equation as 
\begin{equation}
H(\sigma, z) = f_1(\sigma)\,H_1(z) + f_2(\sigma)\,H_2(z).\label{H2hom}
\end{equation}
Here $f_1$ and $f_2$ are the two arbitrary functions (replacing 
$\widetilde{\alpha}$ and $\widetilde{\beta}$) and
\begin{equation}
H_1(z) = {\frac {q_0}{\sqrt{p}}},\qquad H_2(z) = {\frac {\sqrt{\Lambda}\,z} 
{2\sqrt{p}}}.\label{H2g}
\end{equation}
In accordance with our analysis above, the vacuum solution (\ref{H2hom}) 
yields a vanishing 1-form (\ref{gam}), and the curvature of the 3-spacetime 
again reduces to a de Sitter one $R^{\alpha\beta} = -\,\Lambda\,
\vartheta^\alpha\wedge\vartheta^\beta$. 

It is straightforward to take the electromagnetic source into account. 
Integration of (\ref{ddf}) yields the scalar potential
\begin{equation}
\varphi = \nu(\sigma)\varphi_0(z),\qquad \varphi_0(z) = \arctan
(\sqrt{\Lambda}z/2),\label{phi1}
\end{equation}
with an arbitrary function $\nu(\sigma)$. The particular solution of the 
inhomogeneous equation (\ref{ddH}) then reads:
\begin{equation}\label{in1}
H_i(\sigma, z) = {\frac {4\pi G\varepsilon_0\nu^2\Lambda}{c^2(\Lambda\alpha^2
+ \beta^2)\sqrt{p}}}\left[2\sqrt{\Lambda}(\alpha z - \beta q_0/\Lambda)
\,\arctan(\sqrt{\Lambda}z/2) - q\,\ln(p/|q|)\right].
\end{equation}
Correspondingly, for $N=1$, we find the general solution of (\ref{ddH}) as
\begin{equation}\label{N1sol}
H(\sigma, z) = f_1(\sigma)\,H_1(z) + f_2(\sigma)\,H_2(z) + H_i(\sigma, z).
\end{equation}

\subsection{$N=2$: Four-dimensional spacetime}

For $N=2$ it is convenient to combine the two real coordinates $z^a, a=1,2,$
into a complex variable $\zeta = z^1 + iz^2$. Then (\ref{ddH}) can be
recast into
\begin{equation}\label{ddH2}
\partial^2_{\zeta\overline{\zeta}}H + {\frac {\Lambda}{2p^2}}\,H = 
{\frac {16\pi G\varepsilon_0\,p}{c^2\,q}}\,\partial_\zeta\varphi
\partial_{\overline{\zeta}}\varphi. 
\end{equation}
We now have $p = 1 + \Lambda\zeta\overline{\zeta}/4$ and $q =(1 -\Lambda\zeta
\overline{\zeta}/4)\,\alpha + (\zeta\overline{\beta}+\overline{\zeta}\beta)/2$ 
with $\beta = \beta_1 + i\beta_2$. The overbar denotes complex conjugate 
quantities. The integration of the equation (\ref{ddH2}) is based on the 
following differential identity which holds for an arbitrary function 
$f(\sigma, \zeta, \overline{\zeta})$:
\begin{equation}
\partial^2_{\zeta\overline{\zeta}}\left[p^2\partial_\zeta\left({\frac f{p^2}}
\right)\right] + {\frac {\Lambda}{2p^2}}\,\left[p^2\partial_\zeta\left(
{\frac f{p^2}}\right)\right]\equiv\partial_\zeta\left[p^2\partial_\zeta\left(
{\frac {\partial_{\overline{\zeta}} f} {p^2}}\right)\right].\label{id}
\end{equation}
As a result, we have the general solution of (\ref{ddH2}) in the form,
\begin{equation}
H(\sigma, \zeta, \overline{\zeta}) = p^2\left[\partial_\zeta\left({\frac 
f{p^2}}\right) + \partial_{\overline{\zeta}}\left({\frac {\overline{f}}{p^2}}
\right)\right],\qquad f = f_0(\sigma, \zeta) + f_1(\sigma, \zeta, 
\overline{\zeta}),\label{N2sol}
\end{equation}
where $f_0(\sigma, \zeta)$ is an arbitrary holomorphic function of $\zeta$ and
\begin{equation}
f_1(\sigma, \zeta, \overline{\zeta}) = {\frac {8\pi G\varepsilon_0}{c^2}}
\,\int^{\overline{\zeta}} d\overline{\zeta}\,p^2\int^\zeta d\zeta'\,p^{-2} 
\int^{\zeta'}d\zeta''\,(p/q)\partial_{\zeta''}\varphi\partial_{\overline{\zeta}
''}\varphi.\label{f1}
\end{equation}
Equation (\ref{ddf}) reduces to $\partial^2_{\zeta\overline{\zeta}}\varphi =0$
which means that the electromagnetic potential can be expressed as $\varphi =
\varphi_0 + \overline{\varphi}_0$ in terms of an arbitrary holomorphic 
function $\varphi_0(\sigma, \zeta)$.

The 4-dimensional problem was studied in detail in \cite{orr,dirk1,dirk2,pod}, 
and the above solution was derived together with the explicit computation 
of the integral (\ref{f1}).

\subsection{Arbitrary $N$: Rotationally symmetric solutions}

For $N>2$, the complete integration of the partial differential equation
(\ref{ddH}) becomes rather nontrivial. Here we will not consider this
problem in its full generality, but from now on we confine ourselves to
the rotationally symmetric solutions which are described by the functions
$H(\sigma, \xi)$ with the radial variable $\xi = \sqrt{z_a z^a} = 
\sqrt{\delta_{ab}\,z^az^b}$. Recalling (\ref{H1}), we readily have one such 
solution {\it in vacuum} which is given by
\begin{equation}
H_1(\sigma, \xi) = {\frac {q_0}{p^{N/2}}}.\label{H10}
\end{equation}
Note that $p=p(\xi)= 1 + \Lambda\xi^2/4$ and $q_0=q_0(\xi)= 1-\Lambda\xi^2/4$.

The second independent vacuum solution $H_2$ can be obtained by using the 
Liouville formula after rewriting the $N$-dimensional Laplacian of (\ref{ddH}) 
in hyperspherical coordinates: $\partial_a\partial^aH = \xi^{1-N}\partial_\xi
(\xi^{N-1}\partial_\xi H) + \xi^{-2}\Delta_{S_{N-1}}H$ (with $\Delta_{S_{N-1}}$
the Laplacian on a $(N-1)$-hypersphere). Denoting $x=\sqrt{\Lambda}\xi/2$,
we then obtain
\begin{equation}
H_2 = {\frac {q_0}{p^{N/2}}}\int dx\,{\frac {(1 + x^2)^N}{(1 - x^2)^2x^{N-1}}}.
\end{equation}
The final explicit result depends essentially on whether $N$ is an 
odd or an even number.

For the {\it odd} $N= 2n+1, n=0,1,\dots,$ we have the second rotationally
symmetric vacuum solution:
\begin{equation}\label{H2o}
H_2 = 2^{N -2}\,{\frac {q_0} {p^{N/2}}}\,\sum_{k=0}^n {\frac {\left({}_k^n
\right)} {2k-1}}\left({\frac {-\,q_0}{\sqrt{\Lambda}\,\xi}}\right)^{2k-1}.
\end{equation}
For the {\it even} $N= 2n, n=1,2,\dots,$ we find a more complicated result:
\begin{equation}
H_2 = {\frac 1 {4\,p^{N/2}}}\left\{\left({}_{\,\,n}^{2n}
\right)\left[p + q_0n\ln\,(\Lambda\xi^2/4)\right] + \sum_{\stackrel{k\neq 0}
{k=-n}}^{+n}\left[p + q_0n/k\right]\left({}_{n-k}^{\,\,2n}\right)\left(
\Lambda\xi^2/4\right)^k\right\}.\label{H2e}
\end{equation}
As usual, $\left({}_k^n\right)$ denotes the binomial coefficient. It is 
instructive to write down several examples for the lower dimensions. For 
$N=1$ (three-dimensional spacetime), Eq. (\ref{H2o}) yields $H_2=(\sqrt{
\Lambda}\xi/2)/\sqrt{p}$, thus recovering (\ref{H2g}). For $N=2$ (four 
spacetime dimensions), we have {}from (\ref{H2e}): $H_2 = 1 + (q_0/p)\ln
\left(\sqrt{\Lambda}\xi/2\right)$. In a five-dimensional spacetime (for 
$N=3$), the equation (\ref{H2o}) yields $H_2 = 2(p^2 - 2q_0^2)/(\sqrt{
\Lambda}\xi q_0p^{3/2})$. In six dimensions (for $N=4$), we read off from 
(\ref{H2e}): $H_2 = (4/p)\left[1 - q_0^2/(2\Lambda\xi^2) + (3q_0/2p)
\ln\left(\sqrt{\Lambda}\xi/2\right)\right]$. 

The general rotationally symmetric solution {\it in vacuum} reads:
\begin{equation}
H(\sigma, z^a) = f_1(\sigma)\,H_1(z^a) + f_2(\sigma)\,H_2(z^a),
\end{equation}
with $H_{1,2}$ given by (\ref{H10}), (\ref{H2o}), (\ref{H2e}), and the
two arbitrary functions $f_{1,2}(\sigma)$. It is important to notice that 
whereas the function $H$ is rotationally symmetric, the spacetime metric is
not. The frame and the line element involve the functions $q$ and $s$, 
given by (\ref{pq}) and (\ref{pqs}), which are {\it not} rotationally 
symmetric for nontrivial $\beta_a$.

In order to find solutions for the nontrivial matter source, we have
to analyze the electromagnetic field equations. Since we are interested
in the rotationally symmetric configurations, we also impose
this symmetry condition on the electromagnetic field. It is then 
straightforward to verify that a general rotationally symmetric 
solution of (\ref{ddf}) reads (putting $x=\sqrt{\Lambda}\xi/2$):
\begin{equation}
\varphi =  \nu(\sigma)\varphi_0(\xi),\qquad  \varphi_0(\xi) = \int dx
\,{\frac {(1 + x^2)^{N-2}}{x^{N-1}}}.
\end{equation}
The value of this integral essentially depends on $N$. For $N=1$, we 
find (\ref{phi1}), whereas for higher {\it odd} dimensions $N=2n + 1$ 
with $n=1,2,\dots,$ we obtain
 \begin{equation}\label{phi2o}
\varphi_0 = 2^{N -2}\,\sum_{k=0}^{n-1} {\frac {\left(
{}_{\,\,\,\,k}^{n-1}\right)}{2k+1}}\left({\frac {-\,q_0}{\sqrt{\Lambda}
\,\xi}}\right)^{2k+1}.
\end{equation}
For the {\it even} dimensions $N=2n$, with $n=1,2,\dots,$ we find
\begin{equation}
\varphi_0  = {\frac 12}\Bigg[\left({}_{\,\,n-1}^{2n-2}
\right)\ln\,(\Lambda\xi^2/4) + \sum_{\stackrel{k\neq 0}
{k=-n+1}}^{+n-1}{\frac {\left({}_{n-1-k}^{\,\,2n-2}\right)} k}\left(
\Lambda\xi^2/4\right)^k\Bigg].\label{phi2e}
\end{equation}

Now we are in position to find the particular solution of the inhomogeneous
equation (\ref{ddH}). However, we evidently have a problem: the right-hand
side of (\ref{ddH}) contains $q$ which makes the rotationally symmetric
ansatz inconsistent unless we assume that $\beta_a = 0$. Accordingly, we
now put $\beta_a = 0$ and $\alpha = 1$ (which is always possible by the 
rescaling (\ref{trans1}), (\ref{trans2})). Then $q=q_0$ and the direct 
computation then yields the rotationally symmetric solution of the 
inhomogeneous equation (\ref{ddH}):
\begin{equation}
H_i(\sigma,\xi) = {\frac {16\pi G\varepsilon_0\nu^2}{c^2}}\left[\varphi_0
(\xi)\,H_2(\xi) - \psi(\xi)\,H_1(\xi)\right],\label{inhom}
\end{equation}
where (again using the variable $x=\sqrt{\Lambda}\xi/2$)
\begin{eqnarray}
\psi(\xi) &=& \int dx\,{\frac {(1 + x^2)^{N-2}}{x^{N-1}}}\int^x dx'
\,{\frac {(1 + {x'}^2)^N}{(1 - {x'}^2)^2{x'}^{N-1}}}\nonumber\\
&=& {\frac {N-1} 2}\left(\int dx\,{\frac {(1 + x^2)^{N-2}}{x^{N-1}}}\right)^2
- \int dx\,{\frac {(1 + x^2)^{2N-3}}{(1 - x^2)x^{2N-3}}}.
\end{eqnarray}
The last integrals essentially depend on $N$. For $N=1$, we get
\begin{equation}
\psi(\xi) = {\frac 14}\,\ln(p/|q_0|),
\end{equation}
and then (\ref{inhom}) reduces to a particular case of (\ref{in1}) after 
taking into account (\ref{H2g}) and (\ref{phi1}). For higher dimensions, 
$N>1$, we finally find:
\begin{equation}
\psi(\xi) = {\frac {N-1}2}\,\varphi^2_0(\xi) - 2^{2N -4}\left[
\ln\left({\frac {|q_0|}{\sqrt{\Lambda}\xi}}\right) + \sum_{k=1}^{N-2}
\,{\frac 1 {2k}}\left({\frac {p^2}{\Lambda\xi^2}}\right)^k\right].\label{psi1}
\end{equation}

The general rotationally symmetric solution of (\ref{ddH}) for an 
arbitrary $N$ with the electromagnetic wave source thus reads:
\begin{equation}
H(\sigma, \xi) = f_1(\sigma)\,H_1(\xi) + f_2(\sigma)\,H_2(\xi) 
+ H_i(\sigma, \xi).\label{gensolN}
\end{equation}

\section{Negative cosmological constant}\label{Neg}

Above, we have considered the exact solutions for the case of the
positive cosmological constant, $\Lambda > 0$. 
For completeness, let us give the explicit results which corresponds to
the case of the negative cosmological constant, $\Lambda < 0$. It will
be convenient to consider the first lowest values of $N$ separately, 
following the scheme of the previous section.

\subsection{$N = 0$}

Here we have again the geometry (\ref{N0met}) of the two-dimensional spacetime
of the (now, negative) constant curvature with an arbitrary function 
$H(\sigma)$ and the trivial functions $p = 1$ and $q = \alpha(\sigma)$. 

\subsection{$N = 1$}

Instead of (\ref{H2g}) we how find the two independent solution of
the homogeneous equation (\ref{ddH}) to be $H_2(z) = q_0/\sqrt{|p|}$ and 
$H_2(z) = \sqrt{|\Lambda|}z/2\sqrt{|p|}$ with $q_0 = 1 + |\Lambda|z^2/4$,
and the solution of the Maxwell equations now reads
\begin{equation}
\varphi = \nu(\sigma)\varphi_0(z),\qquad \varphi_0(z) = \ln\left({\frac
{1 + \sqrt{|\Lambda|}z/2}{1 - \sqrt{|\Lambda|}z/2}}\right).\label{phi1n}
\end{equation}
This replaces (\ref{phi1}). Then the general solution is again (\ref{N1sol})
with the particular solution (\ref{in1}) of the inhomogeneous equation 
replaced by 
\begin{equation}\label{in1n}
H_i(\sigma, z) = {\frac {4\pi G\varepsilon_0\nu^2\Lambda}{c^2(\Lambda\alpha^2
+ \beta^2)\sqrt{|p|}}}\left[\sqrt{|\Lambda|}(\alpha z - \beta q_0/\Lambda)
\,\ln\left({\frac {1 - \sqrt{|\Lambda|}z/2}{1 + \sqrt{|\Lambda|}z/2}}\right)
- q\,\ln(|p/q|)\right].
\end{equation}

\subsection{$N = 2$}

In the four-dimensional spacetime, the main formulas (\ref{id})-(\ref{f1}) 
remain valid for any sign of $\Lambda$, and the corresponding solutions
with negative cosmological constant are obtained by analytic continuation
from the solutions with positive cosmological constant. See the refs. 
\cite{orr,dirk1,dirk2} where both cases are studied in detail and \cite{pod}
where the complete classification of the solutions is given.

\subsection{Arbitrary $N$}

For the rotationally symmetric solutions with $\Lambda < 0$ and $N > 1$,
we straightforwardly find the following results. We have now $p = 1 - 
|\Lambda|\xi^2/4$ and $q_0 = 1 + |\Lambda|\xi^2/4$. Furthermore, as before, 
the first independent solution of the homogeneous equation (\ref{ddH})
is (\ref{H10}), whereas the particular solution of the inhomogeneous 
equation is again given by (\ref{inhom}), for $\alpha =1, \beta_a =0$. 
However, the previous explicit result (\ref{psi1}) for the function 
$\psi(\xi)$ should be replaced by
\begin{equation}
\psi(\xi) = {\frac {N-1}2}\,\varphi^2_0(\xi) - 2^{2N -4}(-1)^N\left[\ln
\left({\frac {q_0}{\sqrt{|\Lambda|}\xi}}\right) + \sum_{k=1}^{N-2}\,{\frac 
1 {2k}}\left({\frac {-\,p^2}{|\Lambda|\xi^2}}\right)^k\right].\label{psi2}
\end{equation}
Here we assume $N>1$ (the lower dimensional cases are reported in the
previous subsections). The form of the second solution of the homogeneous 
equation $H_2$ and the solution of the Maxwell equation $\varphi_0$ depends 
significantly on whether $N$ is an odd or an even number.

For the {\it odd} $N = 2n +1, n=1,2,\dots$, we obtain the second independent 
homogeneous solution $H_2(\xi)$ and the electromagnetic potential $\varphi_0(
\xi)$, respectively:
\begin{eqnarray}
H_2&=&2^{N -2}\,{\frac {(-1)^nq_0} {p^{N/2}}}\,\sum_{k=0}^n {\frac {(-1)^k
\left({}_k^n\right)} {2k-1}}\left({\frac {q_0}{\sqrt{|\Lambda|}\,\xi}}
\right)^{2k-1},\label{H2on}\\
\varphi_0 &=& 2^{N -2}\,(-1)^{n-1}\,\sum_{k=0}^{n-1} {\frac {(-1)^k
\left({}_{\,\,\,\,k}^{n-1}\right)}{2k+1}}\left({\frac {q_0}{\sqrt{|\Lambda|}
\,\xi}}\right)^{2k+1}.\label{phi2on}
\end{eqnarray}
For the {\it even} dimensions $N=2n$, with $n=1,2,\dots,$ we find more
complicated expressions:
\begin{eqnarray}
H_2 &=& {\frac {(-1)^n} {4\,p^{N/2}}}\left\{\left({}_{\,\,n}^{2n}
\right)\left[p + q_0n\ln\,(|\Lambda|\xi^2/4)\right] + \sum_{\stackrel{k\neq 0}
{k=-n}}^{+n}\left[p + q_0n/k\right]\left({}_{n-k}^{\,\,2n}\right)\left(
-\,|\Lambda|\xi^2/4\right)^k\right\},\label{H2en}\\
\varphi_0  &=& {\frac {(-1)^n} 2}\Bigg[\left({}_{\,\,n-1}^{2n-2}
\right)\ln\,(|\Lambda|\xi^2/4) + \sum_{\stackrel{k\neq 0}{k=-n+1}}^{+n-1}
{\frac {\left({}_{n-1-k}^{\,\,2n-2}\right)} k}\left(
-\,|\Lambda|\xi^2/4\right)^k\Bigg].\label{phi2en}
\end{eqnarray}
The general rotationally symmetric solution of (\ref{ddH}) for an 
arbitrary $N>1$ with the electromagnetic wave source then again is given 
by (\ref{gensolN}), provided we use (\ref{psi2})-(\ref{phi2en}) in it.

\section{Discussion and conclusion}\label{DC}

In this paper, we report on the gravitational wave solutions in any 
spacetime dimension (lower and higher than 4) with an arbitrary cosmological 
constant $\Lambda$. The wave fronts, obtained as the leaves of constant 
$\sigma$, are then the $N$-dimensional surfaces of constant (positive or 
negative, depending on the sign of $\Lambda$) curvature. In four dimensions
(for $N=2$), such solutions have been recently \cite{pod} interpreted as 
the radiative spacetimes with a non-expanding shear-free and twist-free null 
congruence. When $N\neq 2$, neither the Newman-Penrose formalism nor Petrov 
classification is available for the characterization of the new generalized
wave solutions. However, we can formally distinguish several classes of 
solutions according to the values of $\Lambda$ and the combination $\Lambda
\alpha^2 + \beta_a\beta^a$. Following \cite{pod}, we have the generalized
Kundt waves for $\Lambda > 0$, but for $\Lambda < 0$ we find the generalized
$pp$-waves, the Kundt waves, or the ``Lobatchevski planes waves" when $\Lambda
\alpha^2 + \beta_a\beta^a$ is negative, positive, or zero, respectively.
In this paper we have derived the $N$-dimensional generalizations of the
pure gravitational waves of all classes mentioned above, and an extension 
of the Kundt waves for the nontrivial electromagnetic radiation source. 

The case of vanishing $\Lambda=0$ is straightforwardly obtained in our 
general framework. We have then $p=1$ and $q = \alpha + \beta_az^a$. Both, 
the Maxwell equation (\ref{ddf}) and the Einstein equation (\ref{ddH}), 
reduce to the Laplace and the Poisson equation, respectively, in an 
$N$-dimensional Euclidean space. When $\beta_a \neq 0$, the particular
solution reads $H = (16\pi G\varepsilon_0/c^2)(\phi^a\phi_a/\beta_b\beta^b)
\,q\left(\ln\,q - 1\right)$ and $\varphi(\sigma, z^a) = \phi_a(\sigma)z^a$.
A different particular $pp$-wave solution is recovered for $\beta_a=0$:
$H(\sigma, z^a) = h_{ab}(\sigma)z^az^b$ with $h^a{}_a = (8\pi G
\varepsilon_0/c^2)\phi^a\phi_a$. 

Besides the possible applications of the higher-dimensional solutions in
string-theoretic and brane models \cite{str}, the new lower-dimensional wave
solutions appear to be helpful in discussing the problem of the gravitational 
collapse. 

\bigskip
{\bf Acknowledgments}. This work was supported by the Deutsche 
Forschungsgemeinschaft (Bonn) with the grant HE~528/20-1. I thank Friedrich
Hehl for reading of the paper and helpful discussions.

\section{Appendix}\label{app}

Here we collect some details about the connection and the curvature.
The connection 1-form is determined by the first structure equation
$d\vartheta^\alpha + \Gamma_\beta{}^\alpha\wedge\vartheta^\beta =0.$
The explicit solution reads:
\begin{equation}
\Gamma_\alpha{}^\beta = {\frac 12}\left[e_\alpha\rfloor d\vartheta^\beta -
e^\beta\rfloor d\vartheta_\alpha - (e_\alpha\rfloor e^\beta\rfloor 
d\vartheta^\gamma)\,\vartheta_\gamma\right]. 
\end{equation}
Substituting (\ref{cof}), we find the different components of the 
connection as follows:
\begin{eqnarray}
\Gamma_a{}^b &=&\vartheta_a\,\partial^bp -\vartheta^b\,\partial_ap,\label{g1}\\
\Gamma_a{}^B &=& -\,q\,\partial_a\left({p/q}\right)\vartheta^B 
- {\frac {q^2}p}\,(\partial_as)\,\vartheta^{\widehat{0}}\,k^B,\label{g2}\\
\Gamma_A{}^B &=& \left[ (q/p)\,\underline{d}\left({p/q}\right)
- \vartheta^{\widehat{0}}\left(\partial_\rho s - 2q^{-1}\partial_\sigma q
\right)\right]\eta^B{}_A.\label{g3}
\end{eqnarray}
Here $k^A = \delta^A_{\widehat{1}}$ is the null vector introduced above.
The indices $A,B,\dots$ and $a,b,\dots$ are raised and lowered by means of
the metrics (\ref{met}). The 2-dimensional Levi-Civita tensor $\eta_{AB} = 
- \eta_{BA}$ has the components: $\eta^{\widehat{0}}{}_{\widehat{0}} = -1,
\,\eta^{\widehat{1}}{}_{\widehat{1}} = 1$ and $\eta^{\widehat{0}}
{}_{\widehat{1}} = \eta^{\widehat{1}}{}_{\widehat{0}} = 0$. 

The curvature 2-form is constructed from the connection according to
$R_\alpha{}^\beta = d\Gamma_\alpha{}^\beta + \Gamma_\gamma{}^\beta \wedge
\Gamma_\alpha{}^\gamma.$
Substituting here (\ref{g1})-(\ref{g3}), we find explicitly:
\begin{eqnarray}
R_a{}^b &=& p'(-2p/\xi + p')\,\vartheta_a\wedge\vartheta^b 
= -\,\Lambda\,\vartheta_a\wedge\vartheta^b,\label{r1}\\
R_a{}^B &=& -\,{\cal L}_a\wedge\vartheta^B 
+ k^B\,\gamma_a\wedge\vartheta^{\widehat{0}},\label{r2}\\
R_A{}^B &=& 
= -\,{\cal K}\,\vartheta_A\wedge\vartheta^B.\label{r3}
\end{eqnarray}
Here $\xi = \sqrt{z_a z^a} = \sqrt{\delta_{ab}\,z^az^b}$ and the prime $'$ 
denotes the derivative with respect to this variable (note that $p=p(\xi)=
1 + \Lambda\xi^2/4$). Furthermore,
\begin{eqnarray}
{\cal L}_a &=& \underline{D}\left[q\partial_a(p/q)\right] - (q^2/p)
\,\partial_a(p/q)\,\underline{d}(p/q),\\
{\cal K} &=& -\,(p/q)^2\left[\partial^2_{\rho\rho}s 
+ (e_a\rfloor\underline{d}(q/p))^2\right],
\end{eqnarray}
whereas the components of the 1-form $\gamma_a$ are given in (\ref{gam}).
The covariant derivative $\underline{D}$ is defined here by means of the 
connection (\ref{g1}) on an $N$-dimensional space of constant curvature. 
For example, for a covector $v_a$, we have $\underline{D}v_a = \underline{d}
v_a - \Gamma_a{}^b\,v_b$. Finally, taking into account (\ref{pq}) and 
(\ref{pqs}), we can straightforwardly verify that ${\cal L}_a = \Lambda
\vartheta_a$ and ${\cal K} = \Lambda$ for arbitrary $\alpha, \beta_a$. 

In compact form, the components of the curvature 2-form (\ref{r1})-(\ref{r3})
are then given by the equation (\ref{curv}). The Riemann tensor represents 
the components of the curvature 2-form expanded with respect to the coframe:
$R_\alpha{}^\beta = {\frac 12}\,R_{\mu\nu\alpha}{}^\beta\,\vartheta^\mu\wedge
\vartheta^\nu$. Correspondingly, the Ricci tensor $Ric_{\alpha\beta} = 
R_{\mu\alpha\beta}{}^\mu$ represents the components of the 1-form $e_\beta
\rfloor R_\alpha{}^\beta = Ric_{\alpha\beta}\,\vartheta^\beta$. The curvature
scalar is, as usual, $R = e_\alpha\rfloor e_\beta\rfloor R^{\alpha\beta} 
= g^{\alpha\beta}Ric_{\alpha\beta}$.

\end{document}